\documentstyle[aps,preprint]{revtex}
\draft

\begin{document}

\title{
A Kink Soliton Characterizing Traffic Congestion
}

\author{Teruhisa S. Komatsu}
\address{Department of Physics, Tohoku University, Sendai 980-77, Japan}

\author{Shin-ichi Sasa}
\address{Department of Pure and Applied Sciences, 
         College of Arts and Sciences, University of Tokyo, 
         Komaba, Meguro-ku, Tokyo 153, Japan}

\date{\today}
\maketitle

\begin{abstract}
We study traffic congestion by analyzing a one dimensional traffic flow model.
Developing an asymptotic method to investigate the long time behavior near a 
 critical point, we derive
 the modified Korteweg-de Vries (mKdV) equation as the lowest order model.
There are an infinite number of kink solitons to the mKdV equation,
 while it has been found by numerical simulations
 that the kink pattern arising in traffic congestion
 is uniquely determined irrespective of initial conditions.
In order to resolve this selection problem,
 we consider higher order corrections to the mKdV equation and find that 
 there is a kink soliton which can deform continuously
 with the perturbation represented by the addition of these corrections.
With numerical confirmation, we show that
 this continuously deformable kink soliton characterizes traffic congestion.
We also discuss the relationship between traffic congestion
 and slugging phenomenon in granular flow.
\end{abstract}

\pacs{
46.10.+z 
47.54.+r 
47.20.Ky 
}

%
%
\section{Introduction}\label{sec:intro}

%
%
Cooperative behavior in dissipative systems composed of
discrete elements  has attracted considerable interest. In particular, 
granular flow exhibited by powder assemblies under non-equilibrium 
conditions
 \cite{gra1,gra2,gra3,gra4},
traffic flow
 \cite{tra1,tra2,tra3},
 and  collective motion of living organisms
such as birds, fish, insects, bacteria and motional protein
have been investigated  extensively
 \cite{liv1,liv2,liv3,liv4,liv5},
and their spatio-temporal patterns
have been reproduced successfully by computer simulations of
both discrete and continuous models
 \cite{num1,num2,num3,num4,num5,num6,num7,num8}.
These successes suggest that at least some of these phenomena
can be categorized into one universality class.

%
%
As one example, we find  strong resemblance between 
slugging phenomenon in granular flow and 
congestion in traffic flow.   
The problem we address here is to uncover the common structure 
behind such apparently similar phenomena appearing in completely 
different contexts.  We expect that this problem can be studied by 
finding the simplest model which describes the "essential nature" 
of the phenomena.
Toward understanding the slugging phenomenon of granular flow,
we analyzed
phenomenological two fluid models in the previous papers
 \cite{kdvphys1,kdvphys2},
and we derived the K-dV equation by focusing on the marginal point at which 
the uniform fluidized state becomes unstable. Unfortunately, however, 
phenomena which the K-dV equation can describe are quite restricted, and
we were unable to explain slugging phenomenon
even when taking into account higher order terms.
Until now, a simple model describing
slugging phenomenon has never been derived from phenomenological equations.
Study of traffic congestion is in a similar situation.

%
%
One of the reasons for such a poor understanding is that 
many of the phenomenological models to be analyzed are rather complicated.
However, a quite simple model equation exhibiting 
traffic congestion has been recently proposed by Bando et.al.
 \cite{bando1,bando2}.  
In Section \ref{sec:model},
we introduce this model and review the study by Bando et.al.
In Section \ref{sec:reduce}, we analyze this model equation and derive 
the modified K-dV (mKdV) equation at a critical point. 
This equation is universal in the sense that 
it is independent of the materials in question.
However, since the mKdV equation is a completely integrable system, 
almost all solutions (solitons) are structurally unstable, that is, 
they change discontinuously in response to perturbations of
the system. Nevertheless, as shown in Section \ref{sec:DKS},
there are solutions which deform continuously
when the mKdV equation is appended with higher order corrections.
We call such solutions 'deformable solitons',
distinguishing them from the structurally unstable solutions.
We show that there is a unique deformable kink soliton from which,
by continuous deformation,
a kink solution characterizing the essential features
of congestion can be obtained.
In the final section, we discuss several points, including
the correspondence between our study and
a slugging phenomenon in granular flow.


%
%
\section{Model}\label{sec:model}
In the present study, we employ a traffic flow model of the form
\begin{equation}\label{eqn:model}
\ddot{x}_i = a [ U(x_{i+1}-x_i) - \dot{x_i} ]   ,
\end{equation}
which was proposed recently by Bando et al.\cite{bando1,bando2}.
Here, the dot represents differentiation with respect to time,
$i$ is the vehicle index, ordered according to vehicle position
$x_i$ ($1 \le i \le N$),
and $U$ is the velocity which the drivers prefer, depending on
the inter-vehicle distance
 between the car in question and that immediately ahead of it,
$b_i (\equiv x_{i+1}-x_i)$.
Eq.(\ref{eqn:model}) expresses that all drivers tend to
adjust their vehicles' velocities to this velocity $U$
with a constant sensitivity $a$.
The function $U$ is assumed to be monotonically increasing,
with the supplemental conditions that $U(\infty) < \infty$ and $U(0) = 0$.
Although the theoretical analysis developed in this paper
does not depend on the detailed form of $U(b)$,
we assume $U(b)$ takes the form
\begin{equation}\label{eqn:explicitU}
U(b) = \tanh(b-2)+\tanh(2)	,
\end{equation}
when we need to discuss the system behavior concretely.
Further we consider the situation
that the vehicles run on a circuit of a length $L$,
and focus on the ``thermodynamic limit''
that $(L,N) \rightarrow (\infty,\infty)$ with $L/N$ constant.
Then, in the model system under consideration, there are two  parameters,
a sensitivity $a$ and a mean distance $\bar{b} (\equiv L/N)$.

%
%
%
Eq.(\ref{eqn:model}) has a uniform solution written as
\begin{equation}\label{eqn:unisol}
x_i(t)=\bar b i+U(\bar b)t+{\rm constant},
\end{equation}
which expresses that all vehicles run with the same velocity
$U(\bar b)$. Linear stability analysis of this uniform state
was performed \cite{bando1,bando2}, and it was found that the solution
(\ref{eqn:unisol}) is linearly unstable when
\begin{equation}\label{eqn:linear-stability}
        a < 2 U'(\bar{b})       .
\end{equation}
This unstable region in the parameter space $(a, \bar{b})$ is shown in
Fig.\ref{graph:unstable} \cite{bando1,bando2}.
Here, the prime refers to differentiation with respect to $\bar{b}$.
As seen in this graph, there exists a critical value of $a$,
denoted by $a_c$, such that uniform states with any $\bar{b}$ are
linearly stable when $a>a_c$, while uniform states for
some $\bar{b}$ in a neighborhood of $\bar{b}_c$
are unstable when $a<a_c$. In the parameter space,
the point $(a_c, \bar{b}_c)$ is located at the peak of the marginal
stability line defined by the equation $a=2U(\bar{b})$.
We thus can determine the critical point by solving
the equations  $U''(\bar{b}_c)=0$ and $a_c=2U'(\bar{b}_c)$.
For later convenience,
we present the expression of the relevant eigenvalue $\sigma(k)$
 of the $m$-th Fourier mode ($ k\equiv 2\pi m/N $)
 at the critical point $(a_c, \bar{b}_c)$
 in the long wave-length limit ($k\rightarrow 0$):
\begin{equation}\label{eqn:sigma-longwavelength}
\sigma(k) = 
   \frac{a_c}{2} i k
 + \frac{a_c-a}{4}k^2
 - \frac{a_c}{12} i k^3
 - \frac{a_c}{16} k^4		+ O(k^5) .
\end{equation}

%
%
%
System behavior under the condition that the uniform state is unstable  
has been investigated through numerical simulations of
Eq.(\ref{eqn:model}) \cite{bando1,bando2}. 
It has been observed in this case that vehicles separate into several regions
in the circuit, in which vehicles form clusters
 with inter-vehicle distances fixed at the value $b_L$ or $b_H$
 \cite{bando1,bando2}.
Here, $b_L$ and $b_H$ satisfy $b_L < \bar b < b_H$
(see Figs. \ref{graph:unstable-spatio-temporal}
 and \ref{graph:unstable-ini-las-snap}).
Since $U(b_L) < U(b_H)$,
 a cluster with an inter-vehicle distance $b_L$ ($b_H$)
consists of slow (fast) moving vehicles.
It has been interpreted that
a cluster characterized by $b_L$ suffers traffic congestion 
\cite{bando1,bando2}.
We note that each individual vehicle does not remain in a particular cluster.
In fact, vehicles at the edge of clusters behave as follows.
A vehicle at the head of a cluster with inter-vehicle distance $b_H$
encounters a slow-moving vehicle.
Then, the driver decreases his vehicle velocity
as the distance to the preceding vehicle decreases,
and eventually this vehicle becomes a member of a cluster
with inter-vehicle distance $b_L$. 
Inversely, a vehicle at the head of a cluster characterized by $b_L$  
will become a member of a cluster characterized by $b_H$ after some period.
That is, a given vehicle in the circuit
experiences a series of slow- and fast-moving states.
In this way, members of a cluster change in time.

The most remarkable feature is that
the inter-vehicle distances $b_L$ and $b_H$ are determined uniquely,
irrespective of initial conditions, number of clusters, and $\bar{b}$
\cite{bando1,bando2}.
In addition,
there are two kinds of interfaces connecting two neighboring clusters,
that is,
  one connecting the head of a $b_H$ cluster
  to the tail of a $b_L$ cluster,
  and the other connecting the head of a $b_L$ cluster
  to the tail of a $b_H$ cluster.
It was concluded that $b_L$, $b_H$ and the shapes of the interfaces 
 characterize traffic congestion \cite{bando1,bando2}.
In this paper, we regard the interfaces as kink and anti-kink solutions
 of Eq.(\ref{eqn:model}) and develop a theory
 to determine $b_L$, $b_H$ and the shapes of the kink.

%
%
\section{Reduced Equations}\label{sec:reduce}
%
%
When we apply the methods of 
dynamical system theory to Eq.(\ref{eqn:model}),  
it is convenient to rewrite  it in the following form
\begin{eqnarray}
\dot{v}_i &=& a [ U(b_i) - v_i ]	,\label{eqn:model2v}\\
\dot{b}_i &=& v_{i+1} - v_i		,\label{eqn:model2b}
\end{eqnarray}
where $v_i$ corresponds to the $i-$th vehicle velocity.
%
%
We first note that the equality  $\sum_i \dot b_i=0$ holds
 under periodic boundary conditions.
Since this implies that $\sum_i b_i $ is a conserved quantity, 
we  expect that  the  characteristic time  scale of the 
$m$-th Fourier component of the variable $b$ 
goes to infinity  as $k \rightarrow 0$
( see Eq.(\ref{eqn:sigma-longwavelength}) ).
On the contrary, from the form of Eq.(\ref{eqn:model2v}), we  conjecture that
$v_i$ relaxes to some value determined by the configurations of $\{b_i\}$ 
on a time scale of order $1/a$. Thus,
when we are concerned with the long time behavior,
we are allowed to assume that  the value of $v_i$ is determined 
adiabatically by $\{b_i\}$. This statement is expressed by the relation
\begin{equation}\label{eqn:slav}
  v_i(t) = V(b_i, b_{i+1}, b_{i-1},\cdots).
\end{equation}
Substituting Eq.(\ref{eqn:slav}) into 
Eqs.(\ref{eqn:model2v}) and (\ref{eqn:model2b}),
 we obtain the equations for $V$, and 
we can solve these equations by formal polynomial expansions in ${b_i}$.
However, due to the complicated nature of these expressions,
it is difficult to extract the essential nature of solutions.
Thus we will focus on the system behavior near the critical point  
$(a_c,\bar{b}_c)$ to simplify the problem.
With such a simplification,
the nature of kink solutions can be described, as we shall see later.
%
%
Carrying out numerical simulations of Eq.(\ref{eqn:model}),
we found that
 a kink solution appears supercritically below the critical point $a=a_c$.
Therefore we can expect that each of the variables $(b_i,v_i,i,t)$
possesses a self-similar nature at the supercritical bifurcation point.
We seek to determine the scaling properties
of the variables when $a=a_c(1-\epsilon^2)$, where we have introduced 
$\epsilon$ as a small scaling parameter.
{}From the parabolic form of the marginal stability line 
around the critical point, it is expected
that the amplitude of a kink solution
is scaled as $|b-\bar{b}_c| \propto \epsilon$. 
Since the characteristic scales of space and time
diverge at the critical point, using $\sigma(k)$
 given by Eq.(\ref{eqn:sigma-longwavelength}),
we determine the scalings of $i$ and $t$ in the following way.
First, the real part of Eq.(\ref{eqn:sigma-longwavelength}) contains
both a fourth order dissipation term and
 a negative diffusion term whose coefficient is proportional to $a-a_c$.
{}From the balance of these two terms,
$k$ scales in proportion to $\epsilon$.
This leads to the scaling relation $i \propto \epsilon^{-1}$.
Second, the scaling of $t$ is determined
by the lowest order term in Eq.(\ref{eqn:sigma-longwavelength})
, which is given by a dispersion term
proportional to $k^3$ when we eliminate
a propagation term by shifting to a moving coordinate system.
Thus, $t$ is scaled as $t \propto \epsilon^{-3}$.

%
%
Based on these considerations, we express the variable $b_i$ as  
\begin{equation}\label{eqn:scaleb}
b_i(t)  = \bar{b}_c
 + 2 \epsilon \;
\sqrt{U'(\bar{b}_c)/|U'''(\bar{b}_c)|} \;
 \tilde{b}(Z,T)			,
\end{equation}
where $U'(\bar{b}_c) > 0$, $U'''(\bar{b}_c) < 0$,
 and $Z$ and $T$ are a scaled position and time defined by
\begin{eqnarray}
Z  &=& 2 \epsilon ( i - U'(\bar{b}_c) t ) 	,	\label{eqn:scalez}\\
T  &=& \frac{4}{3}  \epsilon^3  U'(\bar{b}_c) t	.	\label{eqn:scalet}
\end{eqnarray}
Here, elaborate prefactors in Eqs.(\ref{eqn:scaleb}),(\ref{eqn:scalez})
and (\ref{eqn:scalet}) are introduced so that we can 
obtain the simplest form of the equation for $\tilde b$.
 (See Eq.(\ref{eqn:mkdv}).)
Then, from Eq.(\ref{eqn:slav}), $v_i$ is expressed as
\begin{equation}\label{eqn:slav-b}
v_i =
  V( \epsilon\,\tilde{b}(Z,T),
     \epsilon\,\tilde{b}(Z+2\epsilon,T),
     \epsilon\,\tilde{b}(Z-2\epsilon,T), \cdots )	.
\end{equation}
Since $\epsilon$ is a small parameter,
a Taylor expansion can be applied to
 the terms like $\tilde{b}(Z+2\epsilon,T)$ in Eq.(\ref{eqn:slav-b}).
This leads to the expression
\begin{equation}\label{eqn:slav-b-e}
v_i = 
\tilde{V}(\epsilon \tilde{b}(Z,T),
	  \epsilon^2\partial_Z\tilde{b}(Z,T),
	  \epsilon^3\partial_Z^2\tilde{b}(Z,T), \cdots )	.
\end{equation}
Substituting Eq.(\ref{eqn:slav-b-e})
 into the right-handed side of Eq.(\ref{eqn:model2b}), we obtain
\begin{eqnarray}
\epsilon^4 \partial_T \tilde{b} &=&
A [ 	\tilde{V}(\epsilon \tilde{b}(Z+2\epsilon,T),
	  \epsilon^2\partial_Z\tilde{b}(Z+2\epsilon,T),
	  \cdots )
-\tilde{V}(\epsilon \tilde{b}(Z,T),
	  \epsilon^2\partial_Z\tilde{b}(Z,T),
	  \cdots )	],
\nonumber \\
&=& \epsilon \partial_Z
\tilde{B}(	\epsilon\tilde{b},
		\epsilon^2\partial_Z\tilde{b},
		\epsilon^3\partial_Z^2\tilde{b},\cdots) ,
\label{eqn:tildeB}
\end{eqnarray}
where $A$ is a constant
 given by $A=(8U'(\bar{b_c})/3)\sqrt{U'(\bar{b}_c)/|U'''(\bar{b}_c)|}$.
Here, since $\tilde{V}$ remains unknown,
 we determine by assuming the following form:
\begin{equation}\label{eqn:expandV}
\tilde{V} = \sum_{n=0}^{\infty} \epsilon^n
 \tilde{V}^{(n)}(\tilde{b},\partial_Z\tilde{b},\partial_Z^2\tilde{b},\cdots) .
\end{equation}
{}From the form of Eq.(\ref{eqn:slav-b-e}),
we find that the terms $\tilde{V}^{(n)}$ are expressed as
\begin{equation}\label{eqn:tildeVn}
\left\{
\begin{array}{rcl}
\tilde{V}^{(0)} &=& \tilde{V}_0					\nonumber,\\
\tilde{V}^{(1)} &=& \tilde{V}_1\tilde{b}			\nonumber,\\
\tilde{V}^{(2)} &=& \tilde{V}_{20}\tilde{b}^2
		  + \tilde{V}_{21}\partial_Z\tilde{b}		\nonumber,\\
\tilde{V}^{(3)} &=& \tilde{V}_{30}\tilde{b}^3
		  + \tilde{V}_{31}\partial_Z\tilde{b}^2
		  + \tilde{V}_{32}\partial_Z^2\tilde{b}		\nonumber,\\
&\cdots& \nonumber,
\end{array}
\right.
\end{equation}
where $\tilde{V}_0,\tilde{V}_1,\tilde{V}_{20},\tilde{V}_{21},
\tilde{V}_{30},\tilde{V}_{31},\tilde{V}_{32}$, etc. are constants
which are calculated in the following way.
First, we substitute
 Eq.(\ref{eqn:expandV}) ( using Eq.(\ref{eqn:tildeVn}) )
 into Eq.(\ref{eqn:model2v}),
where $\dot{v}_i$ is expressed as a functional of $\tilde{b}$,
 using Eqs.(\ref{eqn:slav-b-e}) and (\ref{eqn:tildeB}).
Then, collecting terms of equal order on both sides,
	noting the relation $a=a_c (1-\epsilon^2)$,
 and comparing coefficients of $\tilde{b},\tilde{b}^2,\partial_Z\tilde{b}$ etc.,
successively, we obtain $\tilde{V}^{(n)}$.
( Terms on both sides of the equation are now functionals of $\tilde{b}$. )
Substituting $\tilde{V}$ derived in this way into Eq.(\ref{eqn:tildeB}),
	we get a reduced equation for $\tilde{b}$ in the form:
\begin{equation}\label{eqn:expandB}
\partial_T \tilde{b} = \partial_Z \sum_{n=0} \epsilon^n
B^{(n)}(\tilde{b},\partial_Z\tilde{b},\partial_Z^2\tilde{b},\cdots) .
\end{equation}

Since we know the scalings of ($\tilde{b},Z,T$),
no explicit calculation is necessary to find
which terms can appear in $B^{(0)}$.
Such terms must be of the same order as $\partial_T{\tilde{b}}$,
and therefore the three terms
$\tilde{b}^3, \partial_Z \tilde{b}^2$ and $\partial_Z^2 {\tilde{b}}$
can appear in $B^{(0)}$,
but the term $\partial_Z \tilde{b}^2$ cannot.
This can be explained in the following way.
Since $U(b)-U(\bar{b}_c)$ is an odd function of $b-\bar{b}_c$
 at least up to the order $\epsilon^3$ at the critical point,
$\tilde{V}$ has the same symmetry up to the same order in $\epsilon$.
This implies that $\tilde{V}_{20}$ and $\tilde{V}_{31}$,
 defined in Eq.(\ref{eqn:tildeVn}), vanish.
As a result, $B^{(0)}$ also has the same symmetry at the critical point.
An explicit calculation in fact does yield a
$B^{(0)}$ which contains only $\tilde{b}^3$ and $\partial_Z^2 \tilde{b}$.
Substituting this $B^{(0)}$ into Eq.(\ref{eqn:expandB}),
we finally get the modified K-dV (mKdV) equation,
\begin{equation}\label{eqn:mkdv}
\partial_T \tilde{b}(Z,T)
 = \partial_Z^3 \tilde{b}(Z,T) -  \partial_Z \tilde{b}^{3}(Z,T)	,
\end{equation}
which has been derived in the contexts of
 plasmas, or helical magnet systems, etc.\cite{mkdvphys1,mkdvphys2,mkdvphys3}.

We consider the steady traveling solutions
of the mKdV equation (\ref{eqn:mkdv}).
Substituting $\tilde{b}(Z,T) = g(Z-cT)$ into (\ref{eqn:mkdv}),
we get
\begin{equation}\label{eqn:traveling-eq0}
\frac{{\rm d}^2 g}{{\rm d}Z^2} = {g}^3 - c g + \gamma	,
\end{equation}
where $\gamma$ is an integral constant
which is determined by the boundary conditions.
Since the periodic boundary conditions we assume give $\gamma=0$,
we consider the case $\gamma=0$ in the argument below.
Here, it is worth noting that
the steady solutions of the non-conserved Ginzburg-Landau equation
also satisfy Eq.(\ref{eqn:traveling-eq0}) with $\gamma=0$.
Also, Eq.(\ref{eqn:traveling-eq0}) can be regarded as
the Hamilton equation for a nonlinear oscillator
in a fourth order potential and can be solved analytically.
When $c<0$, the potential is convex everywhere,
 and thus all solutions are unbounded except that one
 which just remains on a peak.
When $c>0$, the potential has double peaks,
 and solutions bounded between these two peaks can exist.
As a result, the following three parameter family of solutions
( with $\gamma=0 )$ are obtained.
\begin{equation}\label{eqn:g0}
g(Z-cT) = g^{(0)}(\theta;m) =
 \sqrt{\frac{2 m c}{1+m}}
 \quad {\rm sn} \left(
 \frac{4 K(m)\,\theta }{2\pi}\;;\; m
 \right)				,
\end{equation}
\begin{equation}
\theta = \Omega^{(0)}(m) (Z-cT) + \phi,\quad
\Omega^{(0)}(m) \equiv \frac{2\pi}{4 K(m)} \sqrt{\frac{c}{1+m}}	,
\end{equation}
where ${\rm sn}(x,m)$ is a Jacobi elliptic function,
	with ${\rm sn}(x,0)=\sin(x) and {\rm sn}(x,1)=\tanh(x)$,
and $K(m)$ is the complete elliptic integral of the first kind \cite{Y.Otsuki}.
This family is parametrized by a traveling velocity $c$ $(c>0)$,
 a modulus $m$ $(0 \le m \le 1)$,
 and a phase $\phi$ $(0 \le \theta < 2\pi)$.
In particular, solutions with $m=1$ take the form of kink patterns. Thus,
there are an infinite number of 
kink solutions parameterized by the traveling velocity $c$
and position $\phi$.

%
%
Now, let us recall that  in  numerical simulations of Eq.(\ref{eqn:model}),
the amplitude of kink solutions is determined uniquely 
irrespective of initial conditions. This leads us to the question:
which solution correspond to the observed one?
For such a selection problem, first of all, the stability of the
solutions must be discussed.   We note here that  the mKdV equation is
a completely integrable system \cite{mkdvmath}. That is,
the solutions are ``neutral'', in the sense that they are specified by
an infinite set of conserved quantities whose values are determined by
the initial conditions. This implies that the stability argument for the
solutions cannot identify the observed one.
Then, it must be noted that integrable systems are
structurally unstable.  If higher order terms
of the equation for $\tilde{b}$ are taken into account, 
the phase space structure changes discontinuously.
We are now led to consider the mKdV equation with higher order corrections.
After some calculation, we obtain
\begin{equation}\label{eqn:mkdv-dissipation}
\dot{\tilde{b}} =
 \partial_Z^3 \tilde{b}
 - \partial_Z \tilde{b}^3
 - \frac{\epsilon}{\alpha} [ \partial_Z^2 
 ( \tilde{b} + \partial_Z^2 \tilde{b} - \alpha \tilde{b}^3 )
 + \beta \partial_Z \tilde{b}^4	] + O(\epsilon^2) ,
\end{equation}
where
$\alpha = 2/3$ and $\beta=0$
when the an explicit form of $U(b)$
given in Eq.(\ref{eqn:explicitU}) is assumed.
Here, $\alpha=2/3$ holds for an arbitrary form of $U(b)$
satisfying $U'''(\bar{b}_c)<0$,
but $\beta$ is derived as $\beta=(U''''(\bar{b}_c)/3)
\sqrt{U'(\bar{b}_c)/|U'''(\bar{b}_c)|^3}$.
In the argument below, 
we consider the general case, in which $\beta \ne 0$.

%
%
\section{deformable kink soliton}\label{sec:DKS}
We now consider how the mKdV steady traveling solutions (\ref{eqn:g0})
are affected by the perturbation resulting in Eq.(\ref{eqn:mkdv-dissipation}).
Substituting $\tilde{b}(Z,T) = g(Z-cT)$	into Eq.({eqn:mkdv-dissipation}),
we obtain the equation for the steady traveling solutions as 
\begin{equation}\label{eqn:traveling-eq}
\frac{{\rm d}^2 g}{{\rm d}Z^2}
 = g^3 - c g + \frac{\epsilon}{\alpha}\left[
\frac{{\rm d}}{{\rm d}Z}\left(
 g + \frac{{\rm d}^2 g}{{\rm d}Z^2} - \alpha g^3 \right)
+ \beta g^4 \right] + \gamma	.
\end{equation}
As before, we will concentrate on the case $\gamma=0$.
%
%
We solve Eq.(\ref{eqn:traveling-eq}) using a perturbation method 
which is similar to the Krylov-Bogoliubov-Mitropolsky method
for nonlinear oscillations \cite{kbm}.
%
%
First, we consider this equation without the perturbation.
Then, since the differential equation for $g$ is second order, 
the solutions can be expressed as trajectories
 in a two dimensional phase space.
For this purpose we must introduce a coordinate system 
in the phase space. While the obvious coordinate system is
 $(g, {\rm d}g/{\rm d}Z)$, we are not limited to this one.
Since we are free to choose coordinate systems, we choose that
which is best suited to treat the present problem.
In the coordinate system we use, $g(Z)$ is expressed as 
\begin{equation}
g(Z)=g^{(0)}(\theta(Z);m(Z)),
\end{equation}
where the function $g^{(0)}$ is given by Eq.(\ref{eqn:g0}).
Using the coordinate system $(\theta, m)$,
we rewrite Eq.(\ref{eqn:traveling-eq0}) as 
\begin{eqnarray}
\frac{{\rm d} \theta}{{\rm d} Z} &=& \Omega^{(0)}(m)	,\\
\frac{{\rm d} m}{{\rm d} Z} &=& 0			.
\end{eqnarray}

%
%
Next, we take the perturbation into account. 
Since a third order derivative term is included in the 
perturbation, the dimension of the phase space becomes three.
However, noting that the motion in the new dimension has 
a small time scale of order $\epsilon$,
 we can treat Eq.(\ref{eqn:traveling-eq})
as a two-dimensional system in  our perturbation theory.
That is,  we express $g$  as 
\begin{equation}\label{eqn:gex}
g=g^{(0)}(\theta;m) + \sum_{n=1}^{\infty}\epsilon^n g^{(n)}(\theta;m),
\end{equation} 
and the evolution of the coordinates $\theta$ and $m$ are written as
\begin{eqnarray}
\frac{{\rm d} \theta}{{\rm d} Z}
 = \Omega
 =& \Omega^{(0)}(m) + 
 &\sum_{n=1}^{\infty}\epsilon^n\, \Omega^{(n)}(m)
				  	,\label{eqn:theta-perturbed}\\
\frac{{\rm d} m}{{\rm d} Z}
 = M
 =&                  
 &\sum_{n=1}^{\infty}\epsilon^n\, M^{(n)}(m)
			     		.\label{eqn:m-perturbed}
\end{eqnarray}
Here, $g^{(n)}$, $\Omega^{(n)}$ and  $M^{(n)}$ can be 
determined by Eq.(\ref{eqn:traveling-eq}) perturbatively. 
We now present the method to calculate 
$g^{(1)}$, $\Omega^{(1)}$ and  $M^{(1)}$.

%
%
Substituting Eq.(\ref{eqn:gex}) into Eq.({\ref{eqn:traveling-eq}) and 
collecting terms of order $\epsilon$, we obtain  
\begin{equation}\label{eqn:Lg1}
\hat{L} g^{(1)} = h^{(1)}, 
\end{equation}
where
\begin{equation}
\hat{L} \equiv - {\Omega^{(0)}}^2 \partial_\theta^2 + 3 {g^{(0)}}^2 - c,
\end{equation}
and 
\begin{equation}
h^{(1)}= \left\{
 2 \Omega^{(0)}
\left(
\Omega^{(1)} \partial_\theta^2 + M^{(1)}\partial_\theta\partial_m
\right) 
+ 
\left(
\partial_m\Omega^{(0)}
\right)
M^{(1)} \partial_\theta
- 
\frac{1}{\alpha}
\left[
\Omega^{(0)}
\left(
 1 + {\Omega^{(0)}}^2 \partial_\theta^2 - 3 \alpha {g^{(0)}}^2
\right)
 \partial_\theta 
+\beta {g^{(0)}}^3
\right]
\right\}
 g^{(0)}								.
\end{equation}
%
%
%
Here, from the fact that $(\theta, m)$ parameterizes the family of solutions
of Eq.(\ref{eqn:traveling-eq0}), we can easily derive the relations
$\hat{L}\cdot \partial_\theta g^{(0)}=0$ and
$\hat{L}\cdot \partial_m g^{(0)}=0$. 
This implies that there is a solution $g^{(1)}$ only under  
a solvability condition. Since we are interested in the case that
the perturbation expansion is well-defined, we enforce the solvability
condition expressed as
\begin{equation}
(\partial_\theta g^{(0)}, h^{(1)})= (\partial_m g^{(0)}, h^{(1)})= 0,
\end{equation}
where $(f,g)=\int_0^{2\pi} d\theta f(\theta)g(\theta)$ for $2\pi$ periodic
functions $f$ and $g$. These two equations yield the expressions of
$\Omega^{(1)}$ and $M^{(1)}$.  Then, $g^{(1)}$ can be obtained
{}from Eq.(\ref{eqn:Lg1}), owing to the solvability condition. 
%
%
We show only the result of $M^{(1)}$  which will be
important in the following discussion. We  obtain
\begin{equation}\label{eqn:t1}
M^{(1)} = \tau(m)[c^{*}(m)-c]	,
\end{equation}
where 
\begin{equation}\label{eqn:tau}
\tau(m) =
\frac{
	\displaystyle\frac{m}{(1+m)^{5/2}}
	\displaystyle\int_0^{4K}\!\!\!\!\!dx
	\left[
		(\partial_x^2 {\rm sn}(x;m))^2
		+ 6m\alpha (\partial_x {\rm sn}(x;m))^2
			   ({\rm sn}(x;m))^2
	\right]
}{	\alpha
	\partial_m
	\left[ \displaystyle\frac{m}{(1+m)^{3/2}}
		\displaystyle\int_0^{4K}\!\!\!\!\!dx
						(\partial_x {\rm sn}(x;m))^2
	\right]
}		,
\end{equation}
and
\begin{equation}\label{eqn:c-star}
c^{*}(m) =
\frac{
(1+m)
\displaystyle\int_0^{4K}\!\!\!\!\!dx (\partial_x {\rm sn}(x;m))^2
}
{
\displaystyle\int_0^{4K}\!\!\!\!\!dx
\left[
(\partial_x^2 {\rm sn}(x;m))^2
+ 6m\alpha (\partial_x {\rm sn}(x;m))^2 ({\rm sn}(x;m))^2
\right]
}		.
\end{equation}
The profiles of the functions $c^{*}(m)$ and $\tau(m)$ are
displayed  in Fig.\ref{graph:c-star-and-tau}.
For later convenience, we define $m^*(c)$ as the inverse 
function of $c^*(m)$. 

%
%
We study the differential equation Eq.(\ref{eqn:m-perturbed})
with Eq.(\ref{eqn:t1})
under the boundary condition $m(Z=0)=m_0$.
We discuss the behavior of solutions for four cases:
(i)   there is no $m^*(c)$ for a given $c$,
(ii)  $m_0=m^*(c)$,
(iii) $m_0<m^*(c)$ and
(iv)  $m_0>m^*(c)$. 
For case (i),
the sign of $M^{(1)}(m(Z))$ is positive (negative) definite for all $Z$.
When $M^{(1)}(m_0)$ is positive (or negative),
$m(Z)$ increases as $Z$ increases (decreases).
Then $m(Z)=1$ is realized at a finite $Z$.
Since $M^{(1)}(m(Z))$ is positive (negative) even at this point,
$m(Z)$ overshoots $1$ as $Z$ increases (decreases) further.
Thus a perturbed solution cannot be defined,
and there is therefore no steady traveling solution in this case.
For case (ii), $M^{(1)}(m_0)=0$.
This leads to the equality $m(Z)=m^*(c)$ for all $Z$, 
that is, the perturbation does not make $m$ inhomogeneous. 
Thus, the expression  
\begin{equation}
g^{(0)}(\theta,m^*(c)) + \epsilon g^{(1)}(\theta,m^*(c)),
\end{equation} 
with $\theta=\Omega (Z-cT)+\phi$ 
gives a steady traveling solution of Eq.(\ref{eqn:mkdv-dissipation}),
where 
\begin{equation}
\Omega=\Omega^{(0)}(m^{*}(c))+\epsilon\Omega^{(1)}(m^{*}(c)).
\end{equation}
We call the unperturbed solution with $(c,m^{*}(c))$
a ``deformable soliton''
because the effects of infinitely small perturbations 
can be absorbed into infinitely small deformations of
the unperturbed solution. The corresponding perturbed solution
is called a ``deformed soliton''.
In case (iii), $M^{(1)}(m_0)$ is negative.
Here, $m(Z)$ increases (decreases) as $Z$ decreases (increases)
 and converges asymptotically to the fixed point $m(Z)=m^{*}(c)$ ($m(Z)=0$).
Thus, we obtain $m(-\infty)=m^{*}(c)$ and $m(\infty)=0$, that is, 
the corresponding solution is a propagating front solution
 connecting the solution characterized by
$(c, m^{*}(c))$ and the uniform state characterized by $m=0$.
In case (iv), $M^{(1)}(m_0)$ is positive.
In this case, $m(Z)$ decreases as $Z$ decreases
 and converges asymptotically to the fixed point $m(Z)=m^{*}(c)$
 similarly to case (iii).
Nevertheless, $m(Z)$ increases as $Z$ increases
 and overshoots $1$ similarly to case (i).
Thus a perturbed solution cannot be defined,
and thus there is no steady traveling solution in this case,
as in case (i).
As a result, 
all solitons except deformable solitons 
are destroyed by the infinitesimal perturbation.
The deformable solitons of the mKdV equation form
a two parameter family parametrized by the modulus $m$
and arbitrary phase $\phi$.
The traveling velocity of the soliton with modulus $m$ 
is given by $c^{*}(m)$.  Note that the soliton with $m=1$ represents
a kink pattern, while the other solitons are periodic.
Therefore, we can expect that the deformed kink soliton
corresponds to that observed
in numerical simulations of the traffic flow model, Eq.(\ref{eqn:model}).

%
%
In order to support this conjecture further, 
we need to check the linear stability of the deformed solitons 
by studying Eq.(\ref{eqn:mkdv-dissipation}). 
However, we have not yet obtained a rigorous result for this problem.
In this paragraph we present a plausible conjecture.
First, one may prove the asymptotic stability of
all kink solutions of the mKdV equation 
by introducing a physically acceptable norm, as 
recently done for  solitary pulse solutions  \cite{pego}.
Then, the deformed kink soliton, 
which is a solution of Eq.(\ref{eqn:mkdv-dissipation}),
can be shown to be stable if it can be proven that the correction
of eigenvalues associated with the linear stability 
is bounded to order $\epsilon$. 
On the other hand, from the fact
 that the tail of the kink solution simply decays exponentially, 
we expect that the kink interacts with a neighboring anti-kink attractively
\cite{kinkatt}.
This implys that periodic deformed solitons
with sufficiently large $m$ are unstable.
Then, since there is no apparent reason to assume that there is a critical $m$ 
below which the stability of these solutions is changed,
we conjecture that all deformed periodic solitons are unstable, while
the deformed kink soliton is stable.

%
%
%
%
%
%
Next we confirm that the deformed kink soliton   
corresponds to that observed
in numerical simulations of Eq.(\ref{eqn:model}).
First, using Eq.(\ref{eqn:g0}), we find that the profile of $b_i$ 
corresponding to the deformable kink soliton is a kink pattern
connecting the two uniform states 
characterized by $b_L$ and $b_H$, where $b_L=\bar b_c-\Delta b$,
and $b_H =\bar b_c+\Delta b$ with 
\begin{equation}\label{eqn:predicted-amplitude}
\Delta b = 2\epsilon\sqrt{\frac{c^*(1) U'(\bar{b}_c)}{|U'''(\bar{b}_c)|}}.
\end{equation}
This expression can be compared with that obtained by
solving the model Eq.(\ref{eqn:model}). 
We thus carried out numerical simulations of Eq.(\ref{eqn:model})
using the fourth order Runge-Kutta method.
In Figs.\ref{graph:kinkamp} and \ref{graph:kinkform},
 the result is shown together with the theoretical curve. 
It is seen that Eq.(\ref{eqn:predicted-amplitude})
gives a good approximation of the numerical result 
for $\epsilon$ smaller than about 0.25.
Although the discrepancy increases as $\epsilon$ becomes
large, we believe that the behavior for larger $\epsilon$ can be 
 accounted for simply by calculating
 higher order terms in Eq.(\ref{eqn:expandB}).
Therefore, we conclude that the kink pattern arising in traffic congestion 
corresponds to the deformed kink soliton.

%
%
\section{Discussion}\label{sec:discussion}

%
%
To this point, we have analyzed Eq.(\ref{eqn:model}) around the critical point 
$(a_c, \bar{b}_c)$.  Similarly, we can obtain the result that system behavior 
near another point on the marginal line
defined by Eq.(\ref{eqn:marginal-stability-line})
is described to lowest order in a perturbative expansion
by the K-dV equation \cite{kdvphys1,kdvphys2,harcri}.
The K-dV equation
is also a completely integrable system with an infinite number of
conserved quantities \cite{kdvmath1,kdvmath2}. Then, owing to the effects
 introduced by higher order terms in the expansion,
 a pulse solution with some amplitude is thought to
 determine the long time behavior \cite{kawa,ei}.
Here, we should remark that this K-dV pulse solution is observed 
only in a quite narrow region in parameter space. 
This contrasts sharply with the case of our deformed kink soliton, 
which is observed in a wider region, even beyond the marginal stability
line. ( This is discussed further below. )
We believe that understanding the difference between these cases is important.
However, we have no theory related to this problem.
We conjecture that a crucial step is to formulate the non-perturbative 
nature of continuous deformations of the solution.

%
%
Further, as shown in Figs.\ref{graph:stable-spatio-temporal} and
\ref{graph:stable-ini-las-snap},
kink patterns  are observed even when the uniform state is stable. 
This result implies that final states depend on initial conditions, 
because the non-uniformity of vehicle velocities vanishes
when the magnitude of fluctuations around the uniform states
are sufficiently small. Moreover, if a kink pattern forms,
its amplitude is determined by only the value of $a$
irrespective of the linear stability of the uniform states.
Then, from Eq.(\ref{eqn:predicted-amplitude}), we can expect that 
the parameter region where kink patterns can appear is approximated well as 
\begin{equation}\label{eqn:kink-appear}
| \bar b -\bar b_c | <
 2 \epsilon \sqrt{\frac{c^{*}(1) U'(\bar{b}_c)}{|U'''(\bar{b}_c)|}} ,
\end{equation}
recalling that $a=a_c(1-\epsilon^2)$. 
The validity of this expression was confirmed by numerical simulations.
On the other hand,
the marginal stability line of uniform states is given by
\begin{equation}\label{eqn:marginal-stability-line}
\bar{b} - \bar{b}_c
 = \pm \epsilon \sqrt{\frac{2U'(\bar{b}_c)}{|U'''(\bar{b}_c)|}} .
\end{equation} 
This line is contained in the region defined by Eq.(\ref{eqn:kink-appear})
because $2c^{*}(1)=5/2$.
Therefore, kink patterns can appear even when the uniform states are stable.

%
%
Finally, we briefly discuss a slugging phenomenon appearing 
in quasi-one dimensional granular flow.
%
%
As a typical experimental system, we consider the sedimentation of 
solid particles (powder) in fluid confined in a narrow pipe. 
%
%
We assume that the density and velocity of solid particles are 
dynamical variables which  can be regarded as a functions of 
the vertical position $z$ owing to
 the system's quasi-one dimensional geometry. 
In order to see the correspondence with the traffic flow model further,
it is convenient to introduce a description employing "powder elements" 
instead of a continuous description. 
That is, the velocity of powder $v(z,t)$
contained in the $i$-th element located at the position $z$
is represented by $v_i(t)$.
(This is nothing but a shift from the Euler picture to the Lagrange picture.)
Let us consider the equation of motion for the powder element. 
The most important force, responsible for the variety of dynamical behavior,
is the drag force due to the fluid around the settling powder element.
This is proportional to the velocity difference between the powder element 
and fluid. Further, it is known that the proportionality coefficient
(drag coefficient) is determined by the powder density.
Taking the drag force and the gravitational force into account,
we obtain a simple form for the equation of motion:
\begin{equation}\label{eqn:sedimentation}
\dot{v}_i =
 - a_s(\rho) v_i - \tilde{g} = a_s(\rho) [ U_s(\rho) - v_i ],
\end{equation}
where $a_s(\rho)$ is a (rescaled) drag coefficient,
$\tilde{g}$ is a specific gravitational acceleration, and
$U_s(\rho) \equiv -\tilde{g}/a_s(\rho)$.
This model can describe a uniform state in which all elements settle with 
the same velocity $U_s(\rho)$. 
There is one unknown function $a_s(\rho)$ in the model, but this 
can be determined experimentally. Actually, from experiments 
on sedimentation of solid particles in three dimensional fluid 
\cite{buscall}, it is known that the sedimentation velocity $U_s$ is
a monotonically decreasing function of the powder density.
Using this experimental data, the drag coefficient $a_s(\rho)$ 
can be determined.

%
%
One may suspect that the model equation (\ref{eqn:sedimentation})
is too simple to describe complicated phenomena
and that for a realistic description we must include 
a dissipation force accounting for collisions
between particles and a chemical potential force.
Still, our model equation (\ref{eqn:sedimentation}) is instructive when 
we consider the correspondence between slugging phenomenon in granular flow 
and traffic congestion, since Eq.(\ref{eqn:sedimentation}) is of
 the same form as the traffic flow model (\ref{eqn:model})
if we neglect the $\rho$ dependence of $a_s$.
(Note that $\rho$ may be related to inter-vehicle distance $b$.)
The problem which needs to be considered is
whether or not the analysis developed above
can be applied to the model system (\ref{eqn:sedimentation}).
The main difference between the two models is the $\rho$ dependence of 
$a_s$, which  causes  the asymmetric term 
$\partial_Z\tilde{b}^2$ in $B^{(0)}$.
 ( Recall the argument appearing before Eq.(\ref{eqn:mkdv}). )
Due to the presence of this term,
 we cannot obtain the mKdV equation
 even when we focus on a critical point.  
Nevertheless, we conjecture that the asymmetric term 
does not change the essential nature of solutions, i.e., the effects 
caused by the asymmetric term can be absorbed into deformation of 
the kink soliton in the mKdV equation. For this reason, we believe that 
slug flow is described by a deformed kink soliton.  
In order to prove this statement, we need to develop a
non-perturbative method describing the continuous deformation of solutions.
To construct  such a theory is a future problem.

%
%
\section{Acknowledgment}
We are grateful to K.Nakanishi for informing us about reference
\cite{bando1,bando2}
and stimulating discussions on the behavior of the model Eq.(\ref{eqn:model}).
We thank M.Bando and K.Hasebe for their encouragement to this work.
We also thank S-I,Ei and T.Ogawa for informing us about reference \cite{pego}
and discussions on the effect of perturbation to integrable systems.
G.C.Paquette is acknowledged for fruitful discussions on the notion
 of structural stability.
This research was supported in part
 by JSPS Research Fellowships for Young Scientists.

%
%

\begin{figure}
\caption{
Marginal stability line of uniform states in parameter space.
The critical point $(a_c,\bar{b}_c)$ is marked by the filled box.
}
\label{graph:unstable}
\end{figure}

\begin{figure}
\caption{
Spatio-temporal pattern exhibited by Eq.(\protect\ref{eqn:model}),
 with $a=1$, $L=70$, and $N=35$.
Positions of vehicles at every 5 time steps are indicated by dots.
The parameters are chosen such that a uniform state is linearly unstable.
}
\label{graph:unstable-spatio-temporal}
\end{figure}

\begin{figure}
\caption{
Profile of vehicle configuration when $t=1000$ (circles with broken line)
with the same parameters as
 in Fig.\protect\ref{graph:unstable-spatio-temporal}.
For the sake of reference, initial conditions (diamonds with solid line)
are also displayed. The horizontal axis $i$ and vertical axis $b_i$
represent the index of vehicles and
the distance between the $i$-th and ($i+1$)-th vehicle, respectively.
}
\label{graph:unstable-ini-las-snap}
\end{figure}

\begin{figure}
\caption{
$c^{*}(m)$ in Eq.(\protect\ref{eqn:c-star}) versus $m$ (solid line) and
$\tau(m)$ in Eq.(\protect\ref{eqn:tau}) versus $m$ (dashed line).
Vertical axes of solid and dashed lines
are graduated on the left and right, respectively.
Note that $c^{*}(1)=5/4$.}
\label{graph:c-star-and-tau}
\end{figure}

\begin{figure}
\caption{
Amplitudes (max$|b-\bar{b}_c|$) of the kink patterns observed
 in numerical simulations are plotted for $\epsilon$ (diamond).
The solid line shows the theoretical curve given by
 Eq.(\protect\ref{eqn:predicted-amplitude}).
}
\label{graph:kinkamp}
\end{figure}

\begin{figure}
\caption{
Profile of kink pattern observed in numerical simulation with
	$\epsilon=2^{-3},N=128$, and $L=256$ (circle).
The solid line shows the deformable kink soliton
	obtained theoretically,
	whose phase ($i$) is adjusted to match that of the numerical one.
}
\label{graph:kinkform}
\end{figure}

\begin{figure}
\caption{
Spatio-temporal pattern exhibited by Eq.(\protect\ref{eqn:model}),
with  $a=1$,$L=70$, and $N=64$.
Positions of vehicles at every 5 time steps are indicated by dots.
The parameters are chosen such that a uniform state is linearly stable.
}
\label{graph:stable-spatio-temporal}
\end{figure}

\begin{figure}
\caption{
Profile of vehicle configuration when $t=1000$ (circles with broken line)
with the same parameters as in Fig.\protect\ref{graph:stable-spatio-temporal}.
For sake of the reference, initial conditions (diamonds with solid line)
are also displayed. The horizontal axis $i$ and vertical axis $b_i$
represent the index of vehicles and
the distance between the $i$-th and ($i+1$)-th vehicle, respectively.
}
\label{graph:stable-ini-las-snap}
\end{figure}

\end{document}